\documentclass[aps,pre,preprint,groupedaddress,
amsmath,amssymb, amsfonts]{revtex4-2}
\usepackage{graphicx}
\usepackage{dcolumn}
\usepackage{bm}
\usepackage[caption=false]{subfig}

\begin{document}


\title{Thermodynamics of chromosome inversions and 100 million years of \textit{Lachancea} evolution}


\author{B K Clark}
\email[]{bkc@ilstu.edu}
\affiliation{Department of Physics, Illinois State University}


\date{\today}

\begin{abstract}
Gene sequences of a deme evolve over time as new chromosome inversions appear in a population via mutations, some of which will replace an existing sequence.  The underlying biochemical processes that generates these and other mutations are governed by the laws of thermodynamics, although the connection between thermodynamics and the generation and propagation of mutations are often neglected. Here, chromosome inversions are modeled as a specific example of mutations in an evolving system. The thermodynamic concepts of chemical potential, energy, and temperature are linked to the input parameters that include inversion rate, recombination loss rate and deme size. An energy barrier to existing gene sequence replacement is a natural consequence of the model. Finally, the model calculations are compared to the observed chromosome inversion distribution of the \textit{Lachancea} genus of yeast. The model introduced in this work should be applicable to other types of mutations in evolving systems.
\end{abstract}


\maketitle


\section{Introduction}

A chromosome inversion, in which a segment of DNA is switched end-for-end within a DNA molecule, is just one way in which a species' DNA is changed.  New species form in part because of these chromosomal rearrangements.  A small sample of examples include Drosophila \cite{noo01, aya05, ran07, yor07}, sunflowers \cite{rie01, rie03}, and primates \cite{lee08, aya05, nav03, nav03a}.  Once a chromosome inversion enters a reproducing population of a species, or deme, it competes with other DNA sequences.  It may replace most or all of the other DNA sequences, disappear after only a generation, or be long-lived within the population.  Chromosome inversions can extend from micro-inversions that are smaller than 100 bp (base pair) \cite{qu20} to inversions large enough to include at least one protein encoding gene in humans \cite{feu10, bou05, pio19}, yeast \cite{vak16}, chickens and mammals \cite{bou05}, and Drosophila \cite{dro07, bhu08, yor07}, for example.   Databases on human genes include gene size \cite{pio19} and inversions and copy number variations \cite{mar14}. Lande \cite{lan79} introduced the first chromosomal rearrangement model by extending Kimura's diffusion based model of genetic drift \cite{kim62}, which was in turn based on the work of Wright and Fisher \cite{fis22, wri31}.  Fixation probabilities for new chromosome sequences have also been calculated \cite{lan79, hed81, spi92, spi98}. More recently, computer simulations have been developed to model chromosome inversions \cite{cla12, cla15} in a deme. In a totally different approach, chromosome dynamics and collisions have been modeled to predict contact frequencies and locations between chromosomes \cite{won12}, which should ultimately provide a connection between chromosome motion in the cellular environment and observed inversions.

The underlying biochemical processes that generate mutations are governed by the laws of thermodynamics, although the connection between thermodynamics and the generation and propagation of inversions is often neglected.  For example, a description of a computational model of the evolutionary dynamics of mutant allele frequencies with migration between two populations \cite{alt11} did not specify how the choice of allele migration rate is related to thermodynamics. The same is true when describing migration between an arbitrary number of populations \cite{lar16}. The exact nature of the mutation was not described in either of these works.  The mutation could be a chromosome inversion or a point mutation.  Clark \cite{cla15} was more specific in describing the evolution of a deme experiencing chromosome inversions, but still did not establish the thermodynamic connection.  

An initial attempt to use thermodynamics to make an evolution related prediction was not very successful. By the time Darwin introduced the theory of evolution \cite{dar59}, thermodynamics was a well established field of physics.  Kelvin estimated the age of the sun using thermodynamics \cite{tom62} and showed a conflict with the minimum age of the earth as required by evolutionary theory.  The conflict was resolved after the sun was shown to undergo fusion and be much older than Kelvin calculated.  Now, physics and more specifically, thermodynamics, contributes to our understanding of DNA. For example, the thermodynamics of the overall organization and topology of DNA \cite{san04,tra13} as well as specific DNA structures including Holliday Junctions \cite{wan16} have been described. 

We discuss a thermodynamics based picture of a deme subject to chromosomal inversions that links the population size, inversion rate and recombination loss rate to the temperature, chemical potential, and inversion energy. Additionally, since the thermodynamics is not specific to the physical characteristics of chromosomal inversions, it should be applicable to other evolutionary processes that can be described with mutation rates and reproduction loss rates. Statistical mechanics concepts are cautiously employed since evolving demes are non-equilibrium systems.

\section{Ensemble Model}

The impact of chromosomal inversions on a deme can be modeled by employing an ensemble description in which every deme in the ensemble has identical initial conditions. Each member of a deme of \textit{n} haploid individuals is modeled as as a single strand of DNA divided into \textit{m} unique protein encoding genes.  The choice to use a haploid population instead of a diploid population as some have done \cite{alt11, lar16, con21} does not affect the general applicability of this model.  Initially the gene sequence is identical for all \textit{n} individuals. Strings of genes move from their original positions as inversions are introduced as time advances. The model only allows two gene sequences to exist in a deme at one time, so new inversions are only introduced into demes with one gene sequence. It is assumed that inversions are introduced at a low enough rate that this is a good approximation. A deme may have between zero and \textit{n} copies of this new sequence as time increases. The distinction between paracentric and pericentric inversions can be made when required. The distinction will only be important in this work when comparing with the \textit{Lachancea} \cite{vak16} results at the end of this article.

The model needs to distinguish between two gene sequence.	The most frequently occurring sequence in a generation is referred to as primary and the less frequent sequence as secondary. All inversions enter the population as secondary sequences. The primary and secondary sequence designations can change with time as the number of copies of each sequence changes. Once an inversion has been introduced into a deme and some time has passed a deme may visit any allowed state, defined as the number of individuals with secondary gene sequences. If the number of secondary sequences exceeds $n/2$, it becomes the primary sequence and the original sequence becomes the secondary sequence. This classification of gene sequences reduces the number of equations required to describe the time evolution of the ensemble from $n + 1$ to $n/2 + 1$. The probability that a deme will be in a state with \textit{i} secondary gene sequences at a time, \textit{t}, is $\rho _{i}  (t)$, $0 \leq i \leq n/2$. 	

Inversion and recombination events can occur each generation. New inversions are introduced at a rate \textit{nI} by transferring the fraction $nI$ of demes in the $i = 0$ state to the $i = 1$ state. If only inversion is included, the process can be described in matrix form as

\begin{equation}		
\begin{pmatrix}
\rho_{0}(t+\Delta t)\\
\rho_{1}(t+\Delta t)\\
\rho_{2}(t+\Delta t)\\
\vdots\\
\rho_{n/2}(t+\Delta t)
\end{pmatrix}
=
\begin{pmatrix}
1-nI & 0 & 0 & \cdots & 0 \\
nI & 1 & 0 & \cdots & 0 \\
0 & 0 & 1 & \cdots & 0 \\
\vdots & \vdots &\vdots & \ddots & \vdots \\
0 & 0 & 0 &\cdots & 1
\end{pmatrix}
\begin{pmatrix}
\rho_{0}(t)\\
\rho_{1}(t)\\
\rho_{2}(t)\\
\vdots\\
\rho_{n/2}(t)
\end{pmatrix}.
\label{eq:one}
\end{equation}

\noindent
Eq.~(\ref{eq:one}) may be written in shorthand notation as 

\begin{equation}
\rho(t+\delta t)= M_{I} \rho(t).
\end{equation}.

\noindent
An unscripted $\rho$ denotes the column vector with length $n/2 + 1$ and ${M_{I}}$ is the $n/2 + 1$ $\times$ $n/2 + 1$ inversion matrix. The only non-zero values are the first two elements of the first column and the remaining diagonal elements.  
  
Recombination can occur between two identical gene sequences or the two different gene sequences. In the latter case the offspring are assumed to die if the recombination locus is in the region of the inversion because neither offspring will have the complete set of \textit{m} genes. The loss of population with each sequence during recombination is $Li(1-i/n)$, where \textit{L} is the recombination loss rate. The value of \textit{L} is just the fractional size of the inversion relative to the length of the total number of $m$ genes with units of per pair.  The same absolute number of individual with each sequence die in recombination, but a greater fraction of the individuals with the less frequent gene sequence is lost. Losses of the two sequences are equal when $i = n/2$.  This confers an effective fitness advantage to the sequence possessed by the greatest number of deme members. The total loss in population from recombination is

\begin{equation}
n_{loss}= 2 L i\left(1-\dfrac{i}{n}\right).
\label{eq:thr}
\end{equation}

The number of offspring with secondary gene sequences, ${n_{s}}$, after recombination is

\begin{equation}
n_{s}= (1-L)i\left(1-\dfrac{i}{n}\right) + (i-1)\left(\dfrac{i}{n}\right),
\label{eq:fou}
\end{equation}		

and the number of offspring with primary gene sequences, ${n_{p}}$,  after recombination is

\begin{equation}
n_{p}= (1-L)i\left(1-\dfrac{i}{n}\right) + (n-i-1)\left(\dfrac{n-i}{n}\right).
\label{eq:fiv}
\end{equation}		

\noindent
Eqs.~(\ref{eq:thr}) through (\ref{eq:fiv}) assume that an individual will not reproduce with itself, and this possibility becomes negligible when $n$ becomes large. Recombination losses to the deme are replaced by scaling the proportion of surviving offspring with each gene sequence so that $n = n_{s} + n_{p}$ and \textit{n} remains constant. The change $\Delta n_{s,i}$ in the size of the population with an initial $i$ secondary gene sequences for an individual deme is
\begin{equation}
 \Delta n_{s,i} = \left(\dfrac{n n_{s}}{n_{p}+n_{s}}\right)-i.
 \label{eq:six}
\end{equation} 

\noindent
As and example, if $-1 \leq \Delta n_{s,i} \leq 0$, then the fraction $-\Delta n_{s,i}$ of demes in state $i$ is transferred to state $(i - 1)$. This is represented in matrix form as

\noindent
\begin{equation}		
\begin{pmatrix}
\rho_{0}(t+\Delta t)\\
\rho_{1}(t+\Delta t)\\
\rho_{2}(t+\Delta t)\\
\vdots\\
\rho_{n/2}(t+\Delta t)
\end{pmatrix}
=
\begin{pmatrix}
1 & -\Delta n_{s,1} & 0 & \cdots & 0 \\
0 & 1 + \Delta n_{s,1} & -\Delta n_{s,2} & \cdots & 0 \\
0 & 0 & 1+ \Delta n_{s,2} & \cdots & 0 \\
\vdots & \vdots & \vdots & \ddots & \vdots \\
0 & 0 & 0 & \cdots & 1
\end{pmatrix}
\begin{pmatrix}
\rho_{0}(t)\\
\rho_{1}(t)\\
\rho_{2}(t)\\
\vdots\\
\rho_{n/2}(t)
\end{pmatrix},
\label{eq:sev}
\end{equation}
\noindent
where inversion and stochastic effects are ignored.
Eq.~(\ref{eq:sev}) may be written in shorthand notation as 
\begin{equation}
\rho(t+\delta t)= M_{L} \rho(t).
\end{equation}.

Recombination has a stochastic component as pairs of individuals within a deme are selected at random to reproduce. Selection probabilities of individuals that possess one of the two gene sequence is described by the binomial distribution. In matrix form,
\begin{equation}
\begin{split}	
\begin{pmatrix}
\rho_{0}(t+\Delta t)\\
\rho_{1}(t+\Delta t)\\
\vdots\\
\rho_{n/2-1}(t+\Delta t)\\
\rho_{n/2}(t+\Delta t)
\end{pmatrix}\\
=
\begin{pmatrix}
b_{0,n,0} & b_{0,n,1} & \cdots & b_{0,n,n/2-1} & b_{0,n,n/2} \\
b_{1,n-1,0} & b_{1,n-1,1} & \cdots & b_{1,n-1,n/2-1} & b_{1,n-1,n/2} \\
\vdots & \vdots & \ddots & \vdots & \vdots \\
b_{n/2-1,n/2+1,0} & b_{n/2-1,n/2+1,1} & \cdots & b_{n/2-1,n/2+1,n/2-1} & b_{n/2-1,n/2+1,n/2} \\
b_{n/2,0} & b_{n/2,1} & \cdots & b_{n/2,n/2-1} & b_{n/2,n/2} \\
\end{pmatrix}
\begin{pmatrix}
\rho_{0}(t)\\
\rho_{1}(t)\\
\vdots\\
\rho_{n/2-1}(t)\\
\rho_{n/2}(t)
\end{pmatrix},
\end{split}
\label{eq:nin}
\end{equation}
  
\noindent
where $b_{i,j,k} = b_{i,k}+b_{j,k}$, $j = n - i$, and 
\begin{equation} 
 b_{i,k} = \dfrac{n!}{i!(n-i)!}  
 \left(\dfrac{k}{n}\right) ^{i} 
 \left(1-\dfrac{k}{n}\right) ^{n-i}
\end{equation} 
\noindent
is the probability of obtaining \textit{i} individuals with the secondary sequence from a deme with \textit{k} individuals with the secondary sequence. So, $b_{0,n,0}$ is always 1 and all other $b_{i,j,0}$ are zero. In general, none of the other $b_{i,j,k}$ are zero. $M_{B}$ includes any changes in the designations of primary and secondary sequences. Eq.~(\ref{eq:nin}) can be written
\begin{equation}
\rho(t+\Delta t)= M_{B} \rho(t),
\end{equation}
\noindent
in abbreviated form.

In an idealized system, the sequence of inversion and recombination might be envisioned as a particular sequence of events, for example

\begin{equation}
\rho(t+\Delta t)= M_{B} M_{L} M_{I} \rho(t).
\label{eq:twlv}
\end{equation}
\noindent

In practice, all individuals do not always reproduce at the same time. Selection affects which individuals experience an inversion and which individuals undergo recombination, but selection is entirely represented in $M_{B}$. A more complete picture might then be better represented as

\begin{equation}
\rho(t+\Delta t)= M_{T} \rho(t),
\label{eq:thrt}
\end{equation}
\noindent
where 

\begin{equation}
M_{T} =  (M_{B} M_{L} M_{I} + M_{B} M_{I} M_{L} + M_{I} M_{B} M_{L} + M_{I} M_{L} M_{B} +
M_{L} M_{I} M_{B} + M_{L} M_{B} M_{I})/6.
\label{eq:frt}
\end{equation}

\noindent
The population densities of the $n/2 + 1$ states should reach steady state so
$\rho(t+\Delta t)= \rho(t),$ and $\rho(t)$ is an eigenvector of $M_{T}$ with an eigenvalue of 1.

The survival rate of chromosomal inversions can be determined within the framework presented here. To this point in the presentation of the model, once an inversion was present in a deme only the number of individuals with the secondary gene sequence has been important. If the survival rate of a new inversion is required, it will be necessary to keep track of when the new gene sequence is the primary or secondary sequence. This is achieved by setting the deme's initial condition to all demes possessing one inverted gene sequence and modifying the model to track the status of the new gene sequence. The terms primary and secondary will keep their meanings, but the original $\rho(t)$ will represent demes in which the majority of members have the original sequence. The density of states for demes in which the majority of members have the new sequence is $\sigma(t)$. 

The matrix $M_{B}$ must be split into two matrices, $M_{Bp}$ and $M_{Bs}$, where \textit{Bp} and \textit{Bs} refers to the  primary and secondary sequences, respectively. Now, 

\begin{equation}
M_{Bp}=
\begin{pmatrix}
b_{0,0} & b_{0,1} & \cdots & b_{0,n/2-1} & b_{0,n/2} \\
b_{1,0} & b_{1,1} & \cdots & b_{1,n/2-1} & b_{1,n/2} \\
\vdots & \vdots & \ddots & \vdots & \vdots \\
b_{n/2-1,0} & b_{n/2-1,1} & \cdots & b_{n/2-1,n/2-1} & b_{n/2-1,n/2} \\
b_{n/2,0} & b_{n/2,1} & \cdots & b_{n/2,n/2-1} & b_{n/2,n/2} \\
\end{pmatrix}
\end{equation} 
\noindent
and
\begin{equation}
M_{Bs}=
\begin{pmatrix}
b_{n,0} & b_{n,1} & \cdots & b_{n,n/2-1} & b_{n,n/2} \\
b_{n-1,0} & b_{n-1,1} & \cdots & b_{n-1,n/2-1} & (b_{n-1,n/2} \\
\vdots & \vdots & \ddots & \vdots & \vdots \\
b_{n/2+1,0} & b_{n/2+1,1} & \cdots & b_{n/2+1,n/2-1} & b_{n/2+1,n/2} \\
b_{n/2,0} & b_{n/2,1} & \cdots & b_{n/2,n/2-1} & b_{n/2,n/2} \\
\end{pmatrix}
\end{equation} 

\noindent
Separating the matrices permits tracking which demes are predominantly composed of individual with the original gene sequence or the new sequence. After a single generation,
\begin{equation}
\rho_{p}(t+\Delta t)= (1/2)((M_{Bp} M_{L}+M_{L} M_{Bp}) \rho(t)+  (M_{Bs} M_{L}+M_{L} M_{Bs}) \sigma(t))
\label{eq:sevt}
\end{equation}
\noindent
and
\begin{equation}
\sigma(t+\Delta t)= (1/2)((M_{Bp}M_{L}+M_{L} M_{Bp}) \sigma(t)+  (M_{Bs}M_{L}+M_{L} M_{Bs}) \rho(t)).
\label{eq:eit}
\end{equation}
\noindent
All demes will eventually be composed of exclusively  original or new sequences, so after sufficient time as elapsed only $\rho_{0}(t)$ and $\sigma_{0}(t)$ have non-zero values. The fraction of chromosome inversions that survive and eliminate the previous gene sequence from the ensemble of demes is just $\sigma_{0}(t)$ as $t$ becomes large.

\section{Results and Discussion}
\subsection{Density of States}
\begin{figure*}
\includegraphics[scale=2.0]{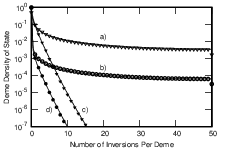}
\caption{\label{fig:fig1}Plot of deme density of states versus number of inversions per deme for cases a) $I=0.001$ (strand generation)$^{-1}$ and $L= 0.00461$ (pair)$^{-1}$, b) $I=0.00001$ (strand generation)$^{-1}$ and $L= 0.00461$ (pair)$^{-1}$, c) $I=0.001$ (strand generation)$^{-1}$ and $L= 0.393$ (pair)$^{-1}$, and d) $I=0.00001$ (strand generation)$^{-1}$ and $L= 0.393$ (pair)$^{-1}$. The deme size is $n = 100$ for all cases. $L= 0.00461$ (pair)$^{-1}$ and $L= 0.393$ (pair)$^{-1}$ correspond to inversion lengths of 4 genes and 256 genes out of total chromosome lengths of 650 genes, respectively. The best fit curves are based on Eq. (\ref{eq:twe}) and overlay the densities calculated with Eq. (\ref{eq:thrt}). Every data point is shown for curves a) and b). Only the points with density of states greater than $10^{-7}$ are shown for curves c) and d). As the density of states curves drop below $10^{-7}$ errors increase until the best fit curves are in error by about a factor of 2 when the density of states have decreased to $10^{-13}$}.
\end{figure*}

The results in this article were obtained using the computer software package Octave \cite{eat19}. The first situation to consider has all demes with the same initial gene sequence and calculates how the demes evolve after many generations.  Four pairs of inversion and recombination loss rates are specified, and the individual element values in $M_{I}$ and $M_{L}$ are determined for each of the four cases. Fig. \ref{fig:fig1} shows typical results for all four initial conditions when the eigenvalues are determined via Eq.~(\ref{eq:thrt}). Nearly identical results are obtained if Eq.~(\ref{eq:thrt}) is solved by computing the column values of $M_{T}^{t}$ where \textit{t} is taken to be a very large number of generations. The density of states of demes with mixed gene sequences is greater for higher inversion and lower recombination loss rates. 

The data in Fig. \ref{fig:fig1} shows that chromosome inversions with lengths that are a significant fraction of the total chromosome length are unlikely to replace the primary gene sequence. For example, the sum of all $\rho_{i}$ for $i \geq 13$ is $1.1\times 10^{-6}$ for sequence inversion length of 256 genes and inversion rate of $1.0 \times 10^{-3}$ (strand generation)$^{-1}$. On average it takes nearly a million generations for the deme to be in one of the states with $i \geq 13$. The fraction of demes with the secondary sequence never becomes large enough to have a significant probability of replacing the primary sequence. Additionally, a deme would have experienced approximately 90000 other chromosome inversions during that million generations, many of which would have occurred within or overlap the bounds of the large inversion. 

The best fit curve for all four sets of data are shown in Fig. \ref{fig:fig1} and are based on the function
\begin{equation}
\rho_i = \frac{g_{i} a}{e^{(\varepsilon i(1-i/n)-\mu)/k_BT}-1},
\label{eq:nint}
\end{equation}			
\noindent	
where the time dependence in the values of $\rho_{i}$ are not included since these values are components of the time independent eigenvector. This is just the Bose-Einstein distribution where \textit{T} is the temperature, $\mu$ is the chemical potential, $g_{i}$ is the degeneracy of the state, and \textit{a} is a normalization constant. The internal energy, $E = \varepsilon i(1-i/n)$, of a deme is proportional to the product of the number of primary and secondary sequences divided by the deme size. A feature of this function of energy is that the energy of state $i$ and state $n-i$ are the same, so $g_{i}=2$ for each state $i$ except $g_{n/2}=1$. The value of $\varepsilon$ can be absorbed into the temperature and chemical potential to yield the dimensionless forms of temperature, $T' = k_{B}T/\varepsilon$, and chemical potential, $\mu'= \mu/\varepsilon$. The $\rho_i$ are then given by

\begin{equation}
\rho_i = \frac{g_{i} a}{e^{(i(1-i/n)-\mu')/T'}-1}.
\label{eq:twe}
\end{equation}			
\noindent
A representation in terms of $\mu'$ and $T'$ is used here. An equally useful representation employs $\mu/(k_{B}T)$ and $\varepsilon/(k_{B}T)$.

The Bose-Einstein \cite{hua63} distribution describes the density of states of some quantum mechanical particles. Normally one would not expect the Bose-Einstein distribution to describe the distribution of states of this non-quantum biological system, but like actual bosons, there is no exclusion principle for demes. There are multiple examples of Bose-Einstein statistics having been successfully applied to biological systems \cite{you87, wil87, jam89}. 

The initial conditions and fitted values of $T'$  and  $\mu'$ for the data in Fig. \ref{fig:fig1} are included in Table~\ref{tab:table1}. The dimensionless form of chemical potential, $\mu'$, is most sensitive to the inversion rate and it's sensitivity can be understood by considering what happens when $T'\gg 1$ for Eqs.~(\ref{eq:twlv}) and (\ref{eq:twe}). Eq.~(\ref{eq:twlv}) represents a simplified version of Eq.~(\ref{eq:frt}) and provides a more tractable view to understand how the thermodynamic variables depend on the input parameters. Since $T'\gg \mu'$, the first few value of $\rho_{i} \approx 2aT'/(i-\mu')$ and the individual values of $a$ and $T'$ are not meaningful. The product $aT'$ becomes the meaningful quantity. The matrix representation in Eq.(12) includes the equation for $\rho_{0}$ in terms of the other $\rho_{i}$. The terms that include $\rho_{i}$ for $i\geq 2$ are small compared to the $\rho_{1}$ term and neglecting these terms yields 
\begin{equation}
\rho_{1}/\rho_{0} \approx \frac{nI(1+\Delta n_{s,1}-b_{0,n,1}(1+\Delta n_{s,1}))}{b_{0,n,1}(1+\Delta n_{s,1})-\Delta n_{s,1}},
\label{eq:ttwo}
\end{equation}	
\noindent
where the linear dependence on $nI$ is explicit. From Eq.~(\ref{eq:twe}), the ratio $\rho_{1}/\rho_{0} = -\mu'$ 
\noindent
for the same condition, so $-\mu'$ has the same linear dependence. The recombination loss term, $\Delta n_{s,1}$,  appears in both the numerator and denominator in a way that makes $\mu'$ not very sensitive to changes in its value. There is no loss from recombination when the the size of the inversion is only one gene because $\Delta n_{s,1} = 0$. The numerator decreases and the denominator increases as the recombination losses increase, since $\Delta n_{s,1}$ becomes more negative. 

\begin{table}[b]
\caption{\label{tab:table1}%
Initial conditions and calculated fit parameters for the data presented in Fig. \ref{fig:fig1}. Notice that $2 a T' = -\mu'$ for $\textit{L}= 0.393$ (pair)$^{-1}$ as expected when $\rho_{0}$ approaches 1. 
}
\begin{ruledtabular}
\begin{tabular}{lcccr}
\textrm{\textit{I}(strand generation)$^{-1}$}&
\textrm{\textit{L}(pair)$^{-1}$}&
\textrm{\textit{T'}}&
\textrm{$\mu'$}&
\textrm{a}\\
\colrule
0.001 & 0.00461 & 47.5 & -0.217 & 0.00117\\
0.00001 & 0.00461 & 71.7 & -0.00191 & 0.0000132\\
0.001 & 0.393 & 0.891 & -0.258 & 0.145\\
0.00001 & 0.393 & 0.902 & -0.00195 & 0.00108\\
\end{tabular}
\end{ruledtabular}
\end{table}

The variation in $\Delta n_{s,1}$ is the basis for the difference between sexual and asexual reproduction in the model. There are no recombination losses when an inverted gene sequence is only one gene in length. There is also no difference between sexual and asexual reproduction since one offspring will be identical to each parent in this case. In asexual reproduction the same similarity between offspring and parents is maintained in every reproduction event, regardless of the size of the inverted gene sequence because there is no recombination or corresponding loss. Every reproduction event leads to viable offspring. In sexual reproduction, recombination losses are allowed and offspring survive in pairs, but with a probability of less than one for inversions including two or more genes. The gene sequence of each parent will be duplicated in one of the offspring if the offspring are viable.  

The observed values for $T'$, $\mu'$, and $\varepsilon$ can be put into perspective by considering the energy required to complete one chromosome inversion.  An inversion requires chemical bonds to be broken between two sets of adjacent nucleic acids on each end of the chromosome inversion and new ones formed.  A barrier to the new genetic sequence must be overcome but the new genetic sequence has nearly the same energy as the old sequence, so very little net energy is required to complete an inversion.  The bond strength of a single deoxyribose nucleotide bond is approximately $2.1 \times 10^{-19}$ Joules \cite{str99}.  Four bonds require approximately $8.4 \times 10^{-19}$ Joules.  

The deme energy as used in Eq.~(\ref{eq:nint}) in an interaction energy. The value can be estimated by assuming the temperature of a biological system is generally near 300 Kelvin.  If we assume this is the temperature, $T$, and take $T'$ to be 100, $\varepsilon$ = $4.1 \times 10^{-23}$ Joules.   If $T'$ decreases to just 4, then $\varepsilon$ increases to $10^{-21}$ Joules. The value of $\varepsilon$ is the maximum amount of energy the deme must gain to replace one individual possessing the primary gene sequence with an individual possessing the secondary sequence. Once the value of the temperature, $T$, is accepted as defined by the environment, the difficulty in determining the value of $T'$ becomes a difficulty in the determining the value of $\varepsilon$ with Eqs.~(\ref{eq:nint}) and ~(\ref{eq:twe}) when $\varepsilon \ll k_{B}T$. In this limit the quantity $a/\varepsilon$ is well defined, but it is still convenient to work with the quantity $aT'$. In all of the cases in Table~\ref{tab:table1}, $-1 < \mu' < 0$, so the chemical potential negligible relative to the available thermal energy. The chemical potential, $\mu$, also becomes more negative as the recombination loss rate increases, but not quite as fast as $\varepsilon$ increases.

The barrier to the replacement of one gene sequence with a new sequence is proportional to the maximum energy of a deme, $\varepsilon (n/4)$. The dimensionless form of the maximum energy of a deme is $n\varepsilon/(4k_{B}T)$. Consequently, the barrier increases as $\varepsilon$ increases or as the deme size increases. Equivalently, the barrier height goes to zero as $\varepsilon$ becomes much less than $k_{B}T$. The parameter values in Table~\ref{tab:table1} show that $T' = k_{B}T/\varepsilon$ is largely dependent on the recombination loss rate. Assuming that $k_{B}T$ is determined by the environment, $\varepsilon$ must increase as the recombination loss rate increases. The increase in the recombination loss rate appears as an increase in the values of the off diagonal elements of $M_{L}$ and a corresponding decrease in the diagonal elements. The deme energy does not have to exceed the barrier energy for the secondary gene sequence to switch roles with the primary sequence, because direct transitions between all states with at least one secondary sequence are possible and given by the elements in $M_{B}$. These transition probabilities are greater for smaller changes in the change in numbers of individuals with each  gene sequence. As the off diagonal elements in $M_{L}$ increase, deme density of states are forced to smaller numbers of secondary sequences so the net transition probabilities from $M_{B}$ acting on $\rho(t)$ that would cause the gene sequences to swap primary and secondary roles are decreased.  

\begin{figure*}
\includegraphics[scale=2.0]{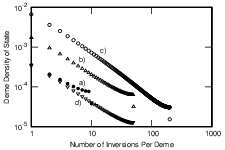}
\caption{\label{fig:fig2}Plot of deme density of states versus number of inversions per deme for cases a) $(\bullet)$ $I=0.00001$ (strand generation)$^{-1}$ and $n = 20$, b) $(\bigtriangleup)$ $I=0.00001$ (strand generation)$^{-1}$ and $n = 100$, c) $(\circ)$ $I=0.00001$ (strand generation)$^{-1}$ and $n = 400$, and d) $(\bigtriangledown)$ $I=0.000002$ (strand generation)$^{-1}$ and $n = 100$. In all cases the inversion size is four genes and the strand length is 650 genes so $L= 0.00461$ (pair)$^{-1}$. The density of states for data sets a) and d) overlap for small numbers of inversions as expected since $nI$ is the same for both sets of data. The density of states curves for b) and d) are parallel for most of the inversions per deme range since the only difference between the two data sets is the inversion rate. In all four cases, the density of states for one inversion per deme are approximately proportional to $nI$. The proportionality breaks down as the number of inversions per deme increases.}
\end{figure*}

The role of population size in the distribution of deme density of states is seen in Fig. \ref{fig:fig2} for deme sizes of $n = $20, 100, and 400 individuals. The density of states for the same value of $i > 0$ is greater for larger deme sizes for constant $I$. This reflects the rate $nI$ that inverted sequences are actually introduced into a deme. If $nI$ is held constant the density of state is a little less for each state $i > 0$ as $n$ increases. This is seen by comparing the density of states for the $n=20$ data with the $n=100$, $I = 2.0 \times 10^{-6}$ (strand generation)$^{-1}$ data.  

\begin{figure*}
\includegraphics[scale=2.0]{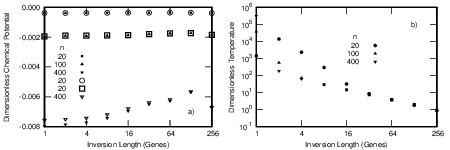}
\caption {\label{fig:fig3} a) Dimensionless chemical potentials calculated for $n=$ 20, 100, and 400 with $I=$ 0.00001 (strand generation)$^{-1}$ are shown as the solid symbols. The values of $-2aT'$ are shown with open symbols and are in good agreement with the values for $\mu'$. The corresponding values for dimensionless temperature are shown in b). The dimensionless temperatures for $n =$ 20 for shorter inversion lengths vary greatly with nearly identical best fit results. However the product $aT'$ is repeatable between best fits as expected for large $T'$.}
\end{figure*}

The fitted values of $\mu'$ and $T'$ for $I = 1.0 \times 10^{-5}$ (strand gen)$^{-1}$ are shown in Fig.~\ref{fig:fig3} as a function of chromosome inversion length. Deme sizes of $n =$ 20, 100, and 400 individuals are included. The values of $\mu'$ do not change much as inversion length changes for each value of $n$. This is expected since the magnitudes of $\mu$ and $\varepsilon$ change together, although not at the same rates. The values of $T'$ decrease as inversion length increases. The data does not extend beyond an inversion length of 256 genes because larger inversions have almost no chance of surviving. The value of $T'$ has already decreased by this point which means the value of $\varepsilon$ has increased to the point that the barrier to gene sequence replacement is large enough to prevent almost all primary gene sequence replacements. Most reported inversions are much shorter than even half the length of a chromosome \cite{feu10, vak16}.  

\subsection{Survival of Chromosomal Inversions}

The discussion presented so far describes the density of states expected for demes with the introduction of new chromosome inversions with a specific length at a rate $I$. Nature presents a species with an evolutionary history that reflects many inversions entering the population over time, some of which survive and others are lost without a trace. The record of surviving inversions available for humans \cite{feu10} is reported in terms of average size of inversions but not directly linked to inversion size in numbers of genes. A chromosome inversion history is available for the \textit{Lachancea} genus of yeast, where numbers of observed chromosome inversions are reported as a function of size in numbers of protein encoding genes \cite{vak16}. 

Yeast reproduces both sexually and asexually, so any model must reflect both modes of reproduction. This is accomplished with the model presented here by modifying $M_{L}$. Now, 

\begin{equation}
M_{L} =(1- X)M_{L_{A}} + XM_{L_{S}}. 
\end{equation}
\noindent
Here, the subscripts $A$ and $S$ reflect asexual and sexual reproduction, respectively. Both components of $M_{L}$ have already been seen. In the asexual case, all $\Delta n_{s,i}$ are set to zero. All $\Delta n_{s,i}$ take on their normal values as already discussed for sexual reproduction. Finally, $X$ reflects the fraction of reproduction that is sexual, so $X = 1$ for all sexual reproduction and $X=0$ for all asexual reproduction.  

\begin{figure*}
\includegraphics[scale=2.0]{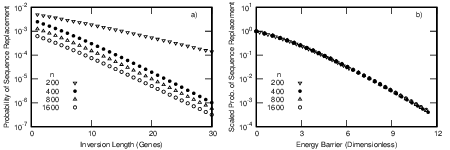}
\caption{\label{fig:fig4}a) is a plot of the probability of an inversion replacing the original gene sequence as a function of inversion length. Results for three different population sizes show the impact of sexual versus asexual reproduction. Data sets $(\bigtriangledown)$ and $(\bullet)$ are for all sexual reproduction and deme sizes of 200 and 400 individuals, respectively. The deme size is increased to 800 individuals for $(\bigtriangleup)$ but only half of the reproduction events are sexual. $(\circ)$ are the results for the largest deme size of 1600 individuals and one quarter of the reproduction events are sexual. b) shows the scaled probability of an inversion replacing the original genes sequencs as a function of the energy barrier, $n'\varepsilon /(k_{B}T)$.}
\end{figure*}

Model results showing the fraction of chromosome inversions that survive and completely replace the original gene sequence are shown in Fig. \ref{fig:fig4}a for inversion sizes up to 30 genes. The figure also shows how asexual and sexual reproduction affect inverted sequence survival. An initial density of states is defined with $\rho_{1} = 1$ and all other $\rho_{i} = 0$. Time increases until essentially every deme is composed of all original gene sequences or all inverted gene sequences. The maximum survival rate is for an inversion length of one gene and is the same for asexual and sexual reproduction. The survival fraction is independent of gene sequence length for completely asexual reproduction so all of the curves for asexual reproduction would be horizontal lines with the same value as shown in the figure for inversions of one gene length. Consequently, only the survival fraction curves that include some sexual reproduction are shown. The four sets of data show the fraction of inversions that survive decreases with increasing inversion length.  The top curve is for $n = 200$ and all sexual reproduction. The other three curves are for $n =$ 400, 800 and 1600 individuals, with sexual reproduction accounting for all, half and one quarter of the reproduction events, respectively. Thus, the number of sexual reproduction events is the same for the last three sets of data.

Fig.~\ref{fig:fig4}a data supports the idea that the probability of a new gene sequence replacing the primary sequence has two components for the inverted gene sequences considered here. First, an inverted chromosome gene sequence in a single individual has a $1/n$ probability of replacing the primary gene sequence for demes without considering the amount of loss due to recombination. This is the value shown for each set of date for inversion lengths of one gene. The second contribution to the probability of sequence replacement is based on the amount of recombination loss. The barrier of dimensionless height $n\varepsilon /(4 k_{B}T)$ that results from recombination loss, reduces the probability of gene sequence replacement by a factor of 
\begin{equation}
e^{-n'\varepsilon /(k_{B}T)},
\label{eq:tthr}
\end{equation}	
\noindent
where $n' = (n/4 -(n-1)/n)$. The last term corrects for the initial state in which all demes have one inverted sequence so the deme has energy $(n-1)\varepsilon/n$. The resulting probability of replacing a primary gene sequence is 
\begin{equation}
P = \frac{e^{-n'\varepsilon /(k_{B}T)}}{n}.
\label{eq:tfo}
\end{equation}	
\noindent
The probability of the new sequence replacing the primary sequence reduces by about 20 \% per each additional gene included in an inversion sequence for sequences with only a few genes and $n$ near 400 for all sexual reproduction. The value of $\varepsilon$ and the resulting barrier to sequence replacement increase as the number of sexual reproduction events increase and as the inversion size increases. The addition of asexual reproduction events decreases the probability of a new sequence surviving primarily through the deme size effect. The results in Fig.~\ref{fig:fig4}a can be plotted as nearly the same function using Eq.~(\ref{eq:tfo}). Fig.~\ref{fig:fig4}b shows the Fig.~\ref{fig:fig4}a data scaled to $nP$ as a function of the argument of the exponential in Eq.~(\ref{eq:tfo}). Data for $n$ as small as 20 nearly overlays the same curve at the resolution of Fig.~\ref{fig:fig4}b, although what appears as a single function in Fig.~\ref{fig:fig4}b is a collection of very closely spaced curves, suggesting Eq.~(\ref{eq:tfo}) is a very good approximation to a more complete function. 

\begin{figure*}
\includegraphics[scale=2.0]{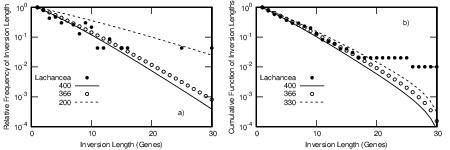}
\caption {\label{fig:fig5} a) Relative frequency of inversions by inversion length for different population sizes. The data for \textit{Lachancea} yeast \cite{vak16} is scaled relative to the number of reported inversions for an inversion size of one gene. The best fit data is for $n=366$ although $n=400$ is visually nearly as good a fit in the figure. The data for $n = 200$ is clearly not a good fit. Lines were used for $n = 200$ and $400$ data to not distract from the yeast and $n= 366$ results. b) Cumulative function of inversion lengths as a function of inversion lengths. The cumulative function begins at large inversion lengths and increases as the inversion length decreases. A least squares function was used to determine that $n = 366$ gives the best fit to the yeast data. The results for $n = 330$ and $400$ represent least squares values that are approximately four times the value for $n = 366$.}
\end{figure*}

The model results can be compared to the observed set of chromosome inversions for the \textit{Lachancea} genus with proper scaling. The number of possible chromosome inversions of a given size depends on the length of the chromosome and whether the inversions are paracentric, pericentric, or include both. The average length of a \textit{Lachancea} chromosome is nearly 647 genes, and the model calculations are based on 650 genes per chromosome. The difference in average length should be inconsequential compared to the factor of two variation in chromosome lengths for \textit{Lachancea}. The observed distribution of chromosome inversions for \textit{Lachancea} is shown in Fig.~\ref{fig:fig5}a as the filled circles. The number of observed inverted gene sequences for lengths up to 30 genes are plotted as a fraction of the number of observed inverted gene sequences with a length of one gene. It should be noted that inversions were observed for all lengths from 1 to 5, 7 to 13, 15, 16, 25 and 30 genes. Beyond inversions of 30 genes in length, there were four reported inversions between 44 and 351 genes in length. The three sets of model results plotted in Fig.~\ref{fig:fig5}a show values at every gene sequence length up to 30 for $n =$ 200, 366, and 400, all sexual reproduction and paracentric inversions. The values are calculated based Eqs.~(\ref{eq:sevt}) and (\ref{eq:eit}) and are scaled by the number of possible inversions of the corresponding gene length. Then they are plotted as a fraction of the number of calculated inversions of a single gene. The model results that best agree with the \textit{Lachancea} data are for $n = 366$. Including pericentric inversions reduces $n$ to 364.

The agreement between the \textit{Lachancea} results and the model calculations for $n = $ 200 is clearly inferior to the agreement with either the $n = $ 366 or 400 data sets. The are several gene sequence lengths where no inversions were observed although they would be expected to exist when comparing to similar gene sequence lengths. A cumulative function for all gene sequence lengths greater than a specific length was used as shown in Fig. \ref{fig:fig5}b, rather than use a data set in a form with multiple zeros. The best least squares fit to the cumulative density function was for $n=$ 366. The continuous curves are for $n =$ 330 and 400, which had 3.8 and 3.9 times the least square error of the $n = $ 366 results.

The same best fit results are obtained if $n = $ 732 or 1464 with one half or one quarter, respectively, of the reproductive events being sexual for the same inversion rate, $I = 10^{-5}$ (strand generation)$^{-1}$. Thus, the model does support a sexually reproductive component, but it does not limit the size of the asexually reproducing component of the population. Referring to Fig. \ref{fig:fig4}, the model does not specify the inversion rate to achieve the observed number of chromosome inversions but shows that a lower inversion rate requires proportionately more generations to achieve the same number of chromosome inversions.

The product of $XIG$ can be estimated, where $X$ and $I$ have already been defined and $G$ is the number of generations per year. \textit{Lachancea} first split into two species about 100 million years ago and now there are ten extant species with chromosome inversion records.\cite{vak16}. Each of the ten species has eight chromosomes. The average time, $t$, each species has existed is about 50 million years. The number of expected inversions in this time period for a specified inversion length, $i$, is 
\begin{equation}
N_{I,i} = Pf_{i}nN_{ch}XIGt. 
\end{equation}
\noindent
The fraction of inversions of gene length $i$ is $f_{i}$. If the calculation is completed for an inversion length of one gene and $n = 366$, $P =0.00273$ and $f_{1} = 0.0062$. $N_{ch}= 80$ is the number of evolving chromosomes. $N_{I,1} = 23$ for \textit{Lachancea}. This gives $XIG = 5.75 \times 10^{-9}$ (strand)$^{-1}$. The value of $X$ is on the order of $1/1000$ to $1/10000$ \cite{tsa08, nie16} and $G$ is between 150 and 2920 (generations/year) \cite{fay05, gal16} for members of the related and better studied genus \textit{Saccharomyces}. The product $XG$ is in the range of 0.150 to 2.92 (generations/year), giving a value of $I$ between $1.22 \times 10^{-11}$ and $2.37 \times 10^{-9}$(strand generation)$^{-1}$.

\section{Conclusion}
The overall model predictions in this work are in good agreement with the historical record of chromosome inversions that have contributed to the differentiation of the ten extant \textit{Lachancea} species of yeast. There are two observations that should be explored further. There is only a 0.0002 chance that no inversions six genes in length would be observed out of the 102 recorded inversions. Advances in experimental measurements and greater numbers of observations will likely resolve this discrepancy. Additionally, the model does not account for the presence of an inversion over 300 genes in length. Catastrophic events are beyond the scope of the neutral model presented here, and could easily account for the success of large inversions. Additional experimental measurements and an expanded model beyond neutral inversions may resolve the differences.  

The chromosome inversions modeled in this article have been limited to neutral inversions. The model can be extended to beneficial, harmful, and underdominant inversions in future work. Then evolving gene sequences of all inversion types can be understood in terms of thermodynamics and the concept of a barrier to gene sequence replacement.

\nocite{*}
\bibliography{thermochominv}

\providecommand{\noopsort}[1]{}\providecommand{\singleletter}[1]{#1}%
\begin{thebibliography}{45}%
\makeatletter
\providecommand \@ifxundefined [1]{%
 \@ifx{#1\undefined}
}%
\providecommand \@ifnum [1]{%
 \ifnum #1\expandafter \@firstoftwo
 \else \expandafter \@secondoftwo
 \fi
}%
\providecommand \@ifx [1]{%
 \ifx #1\expandafter \@firstoftwo
 \else \expandafter \@secondoftwo
 \fi
}%
\providecommand \natexlab [1]{#1}%
\providecommand \enquote  [1]{``#1''}%
\providecommand \bibnamefont  [1]{#1}%
\providecommand \bibfnamefont [1]{#1}%
\providecommand \citenamefont [1]{#1}%
\providecommand \href@noop [0]{\@secondoftwo}%
\providecommand \href [0]{\begingroup \@sanitize@url \@href}%
\providecommand \@href[1]{\@@startlink{#1}\@@href}%
\providecommand \@@href[1]{\endgroup#1\@@endlink}%
\providecommand \@sanitize@url [0]{\catcode `\\12\catcode `\$12\catcode
  `\&12\catcode `\#12\catcode `\^12\catcode `\_12\catcode `\%12\relax}%
\providecommand \@@startlink[1]{}%
\providecommand \@@endlink[0]{}%
\providecommand \url  [0]{\begingroup\@sanitize@url \@url }%
\providecommand \@url [1]{\endgroup\@href {#1}{\urlprefix }}%
\providecommand \urlprefix  [0]{URL }%
\providecommand \Eprint [0]{\href }%
\providecommand \doibase [0]{https://doi.org/}%
\providecommand \selectlanguage [0]{\@gobble}%
\providecommand \bibinfo  [0]{\@secondoftwo}%
\providecommand \bibfield  [0]{\@secondoftwo}%
\providecommand \translation [1]{[#1]}%
\providecommand \BibitemOpen [0]{}%
\providecommand \bibitemStop [0]{}%
\providecommand \bibitemNoStop [0]{.\EOS\space}%
\providecommand \EOS [0]{\spacefactor3000\relax}%
\providecommand \BibitemShut  [1]{\csname bibitem#1\endcsname}%
\let\auto@bib@innerbib\@empty
\bibitem [{\citenamefont {Noor}\ \emph {et~al.}(2001)\citenamefont {Noor},
  \citenamefont {Grams}, \citenamefont {Bertucci},\ and\ \citenamefont
  {Reiland}}]{noo01}%
  \BibitemOpen
  \bibfield  {author} {\bibinfo {author} {\bibfnamefont {M.}~\bibnamefont
  {Noor}}, \bibinfo {author} {\bibfnamefont {K.}~\bibnamefont {Grams}},
  \bibinfo {author} {\bibfnamefont {L.}~\bibnamefont {Bertucci}},\ and\
  \bibinfo {author} {\bibfnamefont {J.}~\bibnamefont {Reiland}},\ }\bibfield
  {title} {\bibinfo {title} {Chromosomal inversions and the reproductive
  isolation of species},\ }\href@noop {} {\bibfield  {journal} {\bibinfo
  {journal} {Proc. Natl. Acad. Sci. U.S.A.}\ }\textbf {\bibinfo {volume}
  {98}},\ \bibinfo {pages} {12084} (\bibinfo {year} {2001})}\BibitemShut
  {NoStop}%
\bibitem [{\citenamefont {Ayala}\ and\ \citenamefont {Coluzzi}(2005)}]{aya05}%
  \BibitemOpen
  \bibfield  {author} {\bibinfo {author} {\bibfnamefont {F.}~\bibnamefont
  {Ayala}}\ and\ \bibinfo {author} {\bibfnamefont {M.}~\bibnamefont
  {Coluzzi}},\ }\bibfield  {title} {\bibinfo {title} {Chromosome speciation:
  Humans, drosophila, and mosquitoes},\ }\href@noop {} {\bibfield  {journal}
  {\bibinfo  {journal} {Proc. Natl. Acad. Sci. U.S.A..}\ }\textbf {\bibinfo
  {volume} {102}},\ \bibinfo {pages} {6535} (\bibinfo {year}
  {2005})}\BibitemShut {NoStop}%
\bibitem [{\citenamefont {Ranz}\ \emph {et~al.}(2007)\citenamefont {Ranz},
  \citenamefont {Maurin}, \citenamefont {Chan}, \citenamefont {von Grotthuss},\
  and\ \citenamefont {Hillier}}]{ran07}%
  \BibitemOpen
  \bibfield  {author} {\bibinfo {author} {\bibfnamefont {J.}~\bibnamefont
  {Ranz}}, \bibinfo {author} {\bibfnamefont {D.}~\bibnamefont {Maurin}},
  \bibinfo {author} {\bibfnamefont {Y.}~\bibnamefont {Chan}}, \bibinfo {author}
  {\bibfnamefont {M.}~\bibnamefont {von Grotthuss}},\ and\ \bibinfo {author}
  {\bibfnamefont {L.}~\bibnamefont {Hillier}},\ }\bibfield  {title} {\bibinfo
  {title} {Principles of genome evolution in the drosophila melanogaster
  species group},\ }\href@noop {} {\bibfield  {journal} {\bibinfo  {journal}
  {PLoS Biology}\ }\textbf {\bibinfo {volume} {5}},\ \bibinfo {pages}
  {doi:10.1371/journal.pbio.0050152} (\bibinfo {year} {2007})}\BibitemShut
  {NoStop}%
\bibitem [{\citenamefont {York}\ \emph {et~al.}(2007)\citenamefont {York},
  \citenamefont {Durrett},\ and\ \citenamefont {Nielsen}}]{yor07}%
  \BibitemOpen
  \bibfield  {author} {\bibinfo {author} {\bibfnamefont {T.}~\bibnamefont
  {York}}, \bibinfo {author} {\bibfnamefont {R.}~\bibnamefont {Durrett}},\ and\
  \bibinfo {author} {\bibfnamefont {R.}~\bibnamefont {Nielsen}},\ }\bibfield
  {title} {\bibinfo {title} {Dependence of paracentric inversion rate on tract
  length},\ }\href@noop {} {\bibfield  {journal} {\bibinfo  {journal} {BMC
  Bioinformatics}\ }\textbf {\bibinfo {volume} {8}},\ \bibinfo {pages}
  {doi:10.1186/1471} (\bibinfo {year} {2007})}\BibitemShut {NoStop}%
\bibitem [{\citenamefont {Rieseberg}(2001)}]{rie01}%
  \BibitemOpen
  \bibfield  {author} {\bibinfo {author} {\bibfnamefont {L.}~\bibnamefont
  {Rieseberg}},\ }\bibfield  {title} {\bibinfo {title} {Chromosomal
  rearrangements and speciation},\ }\href@noop {} {\bibfield  {journal}
  {\bibinfo  {journal} {Trends in Ecol. and Evol.}\ }\textbf {\bibinfo {volume}
  {16}},\ \bibinfo {pages} {351} (\bibinfo {year} {2001})}\BibitemShut
  {NoStop}%
\bibitem [{\citenamefont {Rieseberg}\ \emph {et~al.}(2003)\citenamefont
  {Rieseberg}, \citenamefont {Raymond}, \citenamefont {Rosenthal},
  \citenamefont {Lai}, \citenamefont {Livingston}, \citenamefont {Nakazato},
  \citenamefont {Durphy}, \citenamefont {Schwarzbach}, \citenamefont
  {Donovan},\ and\ \citenamefont {Lexer}}]{rie03}%
  \BibitemOpen
  \bibfield  {author} {\bibinfo {author} {\bibfnamefont {L.}~\bibnamefont
  {Rieseberg}}, \bibinfo {author} {\bibfnamefont {O.}~\bibnamefont {Raymond}},
  \bibinfo {author} {\bibfnamefont {D.}~\bibnamefont {Rosenthal}}, \bibinfo
  {author} {\bibfnamefont {Z.}~\bibnamefont {Lai}}, \bibinfo {author}
  {\bibfnamefont {K.}~\bibnamefont {Livingston}}, \bibinfo {author}
  {\bibfnamefont {T.}~\bibnamefont {Nakazato}}, \bibinfo {author}
  {\bibfnamefont {J.}~\bibnamefont {Durphy}}, \bibinfo {author} {\bibfnamefont
  {A.}~\bibnamefont {Schwarzbach}}, \bibinfo {author} {\bibfnamefont
  {L.}~\bibnamefont {Donovan}},\ and\ \bibinfo {author} {\bibfnamefont
  {C.}~\bibnamefont {Lexer}},\ }\bibfield  {title} {\bibinfo {title} {Major
  ecological transitions in wild sunflowers facilitated by hybridization},\
  }\href@noop {} {\bibfield  {journal} {\bibinfo  {journal} {Science}\ }\textbf
  {\bibinfo {volume} {301}},\ \bibinfo {pages} {1211} (\bibinfo {year}
  {2003})}\BibitemShut {NoStop}%
\bibitem [{\citenamefont {Lee}\ \emph {et~al.}(2008)\citenamefont {Lee},
  \citenamefont {Han}, \citenamefont {Meyer}, \citenamefont {Kim},\ and\
  \citenamefont {Batzer}}]{lee08}%
  \BibitemOpen
  \bibfield  {author} {\bibinfo {author} {\bibfnamefont {J.}~\bibnamefont
  {Lee}}, \bibinfo {author} {\bibfnamefont {K.}~\bibnamefont {Han}}, \bibinfo
  {author} {\bibfnamefont {T.}~\bibnamefont {Meyer}}, \bibinfo {author}
  {\bibfnamefont {H.-S.}\ \bibnamefont {Kim}},\ and\ \bibinfo {author}
  {\bibfnamefont {M.}~\bibnamefont {Batzer}},\ }\bibfield  {title} {\bibinfo
  {title} {Chromosomal inversion between human and chimpanzee lineages caused
  by retrotransposons},\ }\href@noop {} {\bibfield  {journal} {\bibinfo
  {journal} {PLoS ONE}\ }\textbf {\bibinfo {volume} {3}},\ \bibinfo {pages}
  {e4047.doi:10.1371/journal.pone.0004047} (\bibinfo {year}
  {2008})}\BibitemShut {NoStop}%
\bibitem [{\citenamefont {Navarro}\ and\ \citenamefont
  {Barton}(2003{\natexlab{a}})}]{nav03}%
  \BibitemOpen
  \bibfield  {author} {\bibinfo {author} {\bibfnamefont {A.}~\bibnamefont
  {Navarro}}\ and\ \bibinfo {author} {\bibfnamefont {N.}~\bibnamefont
  {Barton}},\ }\bibfield  {title} {\bibinfo {title} {Chromosomal speciation and
  molecular divergence - accelerated evolution in rearranged chromosomes},\
  }\href@noop {} {\bibfield  {journal} {\bibinfo  {journal} {Science}\ }\textbf
  {\bibinfo {volume} {300}},\ \bibinfo {pages} {321} (\bibinfo {year}
  {2003}{\natexlab{a}})}\BibitemShut {NoStop}%
\bibitem [{\citenamefont {Navarro}\ and\ \citenamefont
  {Barton}(2003{\natexlab{b}})}]{nav03a}%
  \BibitemOpen
  \bibfield  {author} {\bibinfo {author} {\bibfnamefont {A.}~\bibnamefont
  {Navarro}}\ and\ \bibinfo {author} {\bibfnamefont {N.}~\bibnamefont
  {Barton}},\ }\bibfield  {title} {\bibinfo {title} {Response to comment on
  chromosomal speciation and molecular divergence - accelerated evolution in
  rearranged chromosomes},\ }\href@noop {} {\bibfield  {journal} {\bibinfo
  {journal} {Science}\ }\textbf {\bibinfo {volume} {302}},\ \bibinfo {pages}
  {988c} (\bibinfo {year} {2003}{\natexlab{b}})}\BibitemShut {NoStop}%
\bibitem [{\citenamefont {Qu}\ \emph {et~al.}(2020)\citenamefont {Qu},
  \citenamefont {Wang}, \citenamefont {He}, \citenamefont {Han}, \citenamefont
  {Yang}, \citenamefont {Wang},\ and\ \citenamefont {Zhu}}]{qu20}%
  \BibitemOpen
  \bibfield  {author} {\bibinfo {author} {\bibfnamefont {L.}~\bibnamefont
  {Qu}}, \bibinfo {author} {\bibfnamefont {L.}~\bibnamefont {Wang}}, \bibinfo
  {author} {\bibfnamefont {F.}~\bibnamefont {He}}, \bibinfo {author}
  {\bibfnamefont {Y.}~\bibnamefont {Han}}, \bibinfo {author} {\bibfnamefont
  {L.}~\bibnamefont {Yang}}, \bibinfo {author} {\bibfnamefont {M.}~\bibnamefont
  {Wang}},\ and\ \bibinfo {author} {\bibfnamefont {H.}~\bibnamefont {Zhu}},\
  }\bibfield  {title} {\bibinfo {title} {The landscape of micro-inversions
  provides clues for population genetic analysis of humans},\ }\href@noop {}
  {\bibfield  {journal} {\bibinfo  {journal} {Interdisciplinary Sciences:
  Computational Life Sciences}\ }\textbf {\bibinfo {volume} {12}},\ \bibinfo
  {pages} {499} (\bibinfo {year} {2020})}\BibitemShut {NoStop}%
\bibitem [{\citenamefont {Feuk}(2010)}]{feu10}%
  \BibitemOpen
  \bibfield  {author} {\bibinfo {author} {\bibfnamefont {L.}~\bibnamefont
  {Feuk}},\ }\bibfield  {title} {\bibinfo {title} {Inversion variants in the
  human genome: role in disease and genome architecture},\ }\href@noop {}
  {\bibfield  {journal} {\bibinfo  {journal} {Genome Medicine}\ }\textbf
  {\bibinfo {volume} {2}},\ \bibinfo {pages} {doi:10.1186/gm132} (\bibinfo
  {year} {2010})}\BibitemShut {NoStop}%
\bibitem [{\citenamefont {Bourque}\ \emph {et~al.}(2005)\citenamefont
  {Bourque}, \citenamefont {Zdobnov}, \citenamefont {Bork},\ and\ \citenamefont
  {Pevzner}}]{bou05}%
  \BibitemOpen
  \bibfield  {author} {\bibinfo {author} {\bibfnamefont {B.}~\bibnamefont
  {Bourque}}, \bibinfo {author} {\bibfnamefont {E.~M.}\ \bibnamefont
  {Zdobnov}}, \bibinfo {author} {\bibfnamefont {P.}~\bibnamefont {Bork}},\ and\
  \bibinfo {author} {\bibfnamefont {P.~A.}\ \bibnamefont {Pevzner}},\
  }\bibfield  {title} {\bibinfo {title} {Comparative architectures of mammalian
  and chicken genomes reveal highly variable rates of genomic rearrangements
  across different lineages},\ }\href@noop {} {\bibfield  {journal} {\bibinfo
  {journal} {Genome Research}\ }\textbf {\bibinfo {volume} {15}},\ \bibinfo
  {pages} {98} (\bibinfo {year} {2005})}\BibitemShut {NoStop}%
\bibitem [{\citenamefont {Piovesan}\ \emph {et~al.}(2019)\citenamefont
  {Piovesan}, \citenamefont {Antonaros}, \citenamefont {Vitale}, \citenamefont
  {Strippoli}, \citenamefont {Pelleri},\ and\ \citenamefont
  {Cracausi}}]{pio19}%
  \BibitemOpen
  \bibfield  {author} {\bibinfo {author} {\bibfnamefont {A.}~\bibnamefont
  {Piovesan}}, \bibinfo {author} {\bibfnamefont {F.}~\bibnamefont {Antonaros}},
  \bibinfo {author} {\bibfnamefont {L.}~\bibnamefont {Vitale}}, \bibinfo
  {author} {\bibfnamefont {P.}~\bibnamefont {Strippoli}}, \bibinfo {author}
  {\bibfnamefont {M.~C.}\ \bibnamefont {Pelleri}},\ and\ \bibinfo {author}
  {\bibfnamefont {M.}~\bibnamefont {Cracausi}},\ }\bibfield  {title} {\bibinfo
  {title} {Human protein-coding genes and gene feature statistics in 2019},\
  }\href@noop {} {\bibfield  {journal} {\bibinfo  {journal} {BMC Res Notes}\
  }\textbf {\bibinfo {volume} {12}},\ \bibinfo {pages} {315} (\bibinfo {year}
  {2019})}\BibitemShut {NoStop}%
\bibitem [{\citenamefont {Vakirlis}\ \emph {et~al.}(2016)\citenamefont
  {Vakirlis}, \citenamefont {Sarilar}, \citenamefont {Drillon}, \citenamefont
  {fleiss}, \citenamefont {Agier}, \citenamefont {Meyniel}, \citenamefont
  {Blanpain}, \citenamefont {Carbone}, \citenamefont {Devillers},\ and\
  \citenamefont {et~al.}}]{vak16}%
  \BibitemOpen
  \bibfield  {author} {\bibinfo {author} {\bibfnamefont {N.}~\bibnamefont
  {Vakirlis}}, \bibinfo {author} {\bibfnamefont {V.}~\bibnamefont {Sarilar}},
  \bibinfo {author} {\bibfnamefont {G.}~\bibnamefont {Drillon}}, \bibinfo
  {author} {\bibfnamefont {A.}~\bibnamefont {fleiss}}, \bibinfo {author}
  {\bibfnamefont {N.}~\bibnamefont {Agier}}, \bibinfo {author} {\bibfnamefont
  {J.}~\bibnamefont {Meyniel}}, \bibinfo {author} {\bibfnamefont
  {L.}~\bibnamefont {Blanpain}}, \bibinfo {author} {\bibfnamefont
  {A.}~\bibnamefont {Carbone}}, \bibinfo {author} {\bibfnamefont
  {H.}~\bibnamefont {Devillers}},\ and\ \bibinfo {author} {\bibfnamefont
  {K.~D.}\ \bibnamefont {et~al.}},\ }\bibfield  {title} {\bibinfo {title}
  {Reconstruction of ancestral chromosome architecture and gene repertoire
  reveals principles of genome evolution in a model yeast genus},\ }\href@noop
  {} {\bibfield  {journal} {\bibinfo  {journal} {Genome Research}\ }\textbf
  {\bibinfo {volume} {41}},\ \bibinfo {pages} {918} (\bibinfo {year}
  {2016})}\BibitemShut {NoStop}%
\bibitem [{\citenamefont {{Drosophila 12 Genomes Consortium}}(2007)}]{dro07}%
  \BibitemOpen
  \bibfield  {author} {\bibinfo {author} {\bibnamefont {{Drosophila 12 Genomes
  Consortium}}},\ }\bibfield  {title} {\bibinfo {title} {Evolution of genes and
  genomes on the drosophila phylogeny},\ }\href@noop {} {\bibfield  {journal}
  {\bibinfo  {journal} {Nature}\ }\textbf {\bibinfo {volume} {450}},\ \bibinfo
  {pages} {203} (\bibinfo {year} {2007})}\BibitemShut {NoStop}%
\bibitem [{\citenamefont {Bhutkar}\ \emph {et~al.}(2008)\citenamefont
  {Bhutkar}, \citenamefont {Schaeffer}, \citenamefont {m.~Russo}, \citenamefont
  {Xu}, \citenamefont {Smith},\ and\ \citenamefont {Gelbart}}]{bhu08}%
  \BibitemOpen
  \bibfield  {author} {\bibinfo {author} {\bibfnamefont {A.}~\bibnamefont
  {Bhutkar}}, \bibinfo {author} {\bibfnamefont {S.~W.}\ \bibnamefont
  {Schaeffer}}, \bibinfo {author} {\bibfnamefont {S.}~\bibnamefont {m.~Russo}},
  \bibinfo {author} {\bibfnamefont {M.}~\bibnamefont {Xu}}, \bibinfo {author}
  {\bibfnamefont {T.~F.}\ \bibnamefont {Smith}},\ and\ \bibinfo {author}
  {\bibfnamefont {M.}~\bibnamefont {Gelbart}},\ }\bibfield  {title} {\bibinfo
  {title} {Chromosomal rearrangement inferred from comparisons of 12 drosophila
  genomes},\ }\href@noop {} {\bibfield  {journal} {\bibinfo  {journal}
  {Genetics}\ }\textbf {\bibinfo {volume} {107}},\ \bibinfo {pages} {1657}
  (\bibinfo {year} {2008})}\BibitemShut {NoStop}%
\bibitem [{\citenamefont {Martinez-Fundichely}\ \emph
  {et~al.}(2014)\citenamefont {Martinez-Fundichely}, \citenamefont {Casillas},
  \citenamefont {Egea}, \citenamefont {R\`{a}mia}, \citenamefont {Barbadilla},
  \citenamefont {Pantano}, \citenamefont {Puig},\ and\ \citenamefont
  {C\'{a}ceres}}]{mar14}%
  \BibitemOpen
  \bibfield  {author} {\bibinfo {author} {\bibfnamefont {A.}~\bibnamefont
  {Martinez-Fundichely}}, \bibinfo {author} {\bibfnamefont {S.}~\bibnamefont
  {Casillas}}, \bibinfo {author} {\bibfnamefont {R.}~\bibnamefont {Egea}},
  \bibinfo {author} {\bibfnamefont {M.}~\bibnamefont {R\`{a}mia}}, \bibinfo
  {author} {\bibfnamefont {A.}~\bibnamefont {Barbadilla}}, \bibinfo {author}
  {\bibfnamefont {L.}~\bibnamefont {Pantano}}, \bibinfo {author} {\bibfnamefont
  {M.}~\bibnamefont {Puig}},\ and\ \bibinfo {author} {\bibfnamefont
  {M.}~\bibnamefont {C\'{a}ceres}},\ }\bibfield  {title} {\bibinfo {title}
  {Invfest, a database integrating information of polymorphic inversions in the
  human genome},\ }\href@noop {} {\bibfield  {journal} {\bibinfo  {journal}
  {Nucleic Acids Res.}\ }\textbf {\bibinfo {volume} {42}},\ \bibinfo {pages}
  {D1027} (\bibinfo {year} {2014})}\BibitemShut {NoStop}%
\bibitem [{\citenamefont {Lande}(1979)}]{lan79}%
  \BibitemOpen
  \bibfield  {author} {\bibinfo {author} {\bibfnamefont {R.}~\bibnamefont
  {Lande}},\ }\bibfield  {title} {\bibinfo {title} {Effective deme sizes during
  long-term evolution estimated from rates of chromosomal rearrangement},\
  }\href@noop {} {\bibfield  {journal} {\bibinfo  {journal} {Evol.}\ }\textbf
  {\bibinfo {volume} {33}},\ \bibinfo {pages} {234} (\bibinfo {year}
  {1979})}\BibitemShut {NoStop}%
\bibitem [{\citenamefont {Kimura}(1962)}]{kim62}%
  \BibitemOpen
  \bibfield  {author} {\bibinfo {author} {\bibfnamefont {M.}~\bibnamefont
  {Kimura}},\ }\bibfield  {title} {\bibinfo {title} {On the probability of
  fixation of mutant genes in a population},\ }\href@noop {} {\bibfield
  {journal} {\bibinfo  {journal} {Genetics}\ }\textbf {\bibinfo {volume}
  {47}},\ \bibinfo {pages} {713} (\bibinfo {year} {1962})}\BibitemShut
  {NoStop}%
\bibitem [{\citenamefont {Fisher}(1922)}]{fis22}%
  \BibitemOpen
  \bibfield  {author} {\bibinfo {author} {\bibfnamefont {R.}~\bibnamefont
  {Fisher}},\ }\bibfield  {title} {\bibinfo {title} {On the dominance ratio},\
  }\href@noop {} {\bibfield  {journal} {\bibinfo  {journal} {Proc. Roy. Soc.
  Edinburgh}\ }\textbf {\bibinfo {volume} {42}},\ \bibinfo {pages} {321}
  (\bibinfo {year} {1922})}\BibitemShut {NoStop}%
\bibitem [{\citenamefont {Wright}(1931)}]{wri31}%
  \BibitemOpen
  \bibfield  {author} {\bibinfo {author} {\bibfnamefont {S.}~\bibnamefont
  {Wright}},\ }\bibfield  {title} {\bibinfo {title} {Evolution in mendelian
  populations},\ }\href@noop {} {\bibfield  {journal} {\bibinfo  {journal}
  {Genetics}\ }\textbf {\bibinfo {volume} {16}},\ \bibinfo {pages} {97}
  (\bibinfo {year} {1931})}\BibitemShut {NoStop}%
\bibitem [{\citenamefont {Hedrick}(1981)}]{hed81}%
  \BibitemOpen
  \bibfield  {author} {\bibinfo {author} {\bibfnamefont {P.}~\bibnamefont
  {Hedrick}},\ }\bibfield  {title} {\bibinfo {title} {The establishment of
  chromosomal variants},\ }\href@noop {} {\bibfield  {journal} {\bibinfo
  {journal} {Evolution}\ }\textbf {\bibinfo {volume} {35}},\ \bibinfo {pages}
  {322} (\bibinfo {year} {1981})}\BibitemShut {NoStop}%
\bibitem [{\citenamefont {Spirito}(1992)}]{spi92}%
  \BibitemOpen
  \bibfield  {author} {\bibinfo {author} {\bibfnamefont {F.}~\bibnamefont
  {Spirito}},\ }\bibfield  {title} {\bibinfo {title} {The exact values of the
  probability of fixation of underdominant chromosomal rearrangements},\
  }\href@noop {} {\bibfield  {journal} {\bibinfo  {journal} {Theo. Pop. Biol.}\
  }\textbf {\bibinfo {volume} {41}},\ \bibinfo {pages} {111} (\bibinfo {year}
  {1992})}\BibitemShut {NoStop}%
\bibitem [{\citenamefont {Spirito}(1994)}]{spi98}%
  \BibitemOpen
  \bibfield  {author} {\bibinfo {author} {\bibfnamefont {F.}~\bibnamefont
  {Spirito}},\ }in\ \href@noop {} {\emph {\bibinfo {booktitle} {Endless Forms:
  Species and Speciation}}},\ \bibinfo {editor} {edited by\ \bibinfo {editor}
  {\bibfnamefont {D.}~\bibnamefont {Howard}}\ and\ \bibinfo {editor}
  {\bibfnamefont {S.}~\bibnamefont {Berlocher}}}\ (\bibinfo  {publisher}
  {Oxford University Press},\ \bibinfo {address} {Oxford},\ \bibinfo {year}
  {1994})\ pp.\ \bibinfo {pages} {320--329}\BibitemShut {NoStop}%
\bibitem [{\citenamefont {Clark}\ \emph {et~al.}(2012)\citenamefont {Clark},
  \citenamefont {Wabick},\ and\ \citenamefont {Weidner}}]{cla12}%
  \BibitemOpen
  \bibfield  {author} {\bibinfo {author} {\bibfnamefont {B.}~\bibnamefont
  {Clark}}, \bibinfo {author} {\bibfnamefont {K.}~\bibnamefont {Wabick}},\ and\
  \bibinfo {author} {\bibfnamefont {J.}~\bibnamefont {Weidner}},\ }\bibfield
  {title} {\bibinfo {title} {Inversion and crossover recombination
  contributions to the spacing between two functionally linked genes},\
  }\href@noop {} {\bibfield  {journal} {\bibinfo  {journal} {BioSystems}\
  }\textbf {\bibinfo {volume} {109}},\ \bibinfo {pages} {169} (\bibinfo {year}
  {2012})}\BibitemShut {NoStop}%
\bibitem [{\citenamefont {Clark}(2015)}]{cla15}%
  \BibitemOpen
  \bibfield  {author} {\bibinfo {author} {\bibfnamefont {B.}~\bibnamefont
  {Clark}},\ }\bibfield  {title} {\bibinfo {title} {Modeling the evolution and
  expected lifetime of a deme's principal gene sequence},\ }\href@noop {}
  {\bibfield  {journal} {\bibinfo  {journal} {Let in Biomath}\ }\textbf
  {\bibinfo {volume} {2}},\ \bibinfo {pages} {13} (\bibinfo {year}
  {2015})}\BibitemShut {NoStop}%
\bibitem [{\citenamefont {Wong}\ \emph {et~al.}(2012)\citenamefont {Wong},
  \citenamefont {Marie-Nelly}, \citenamefont {Herbert}, \citenamefont
  {Carrivain}, \citenamefont {Blanc}, \citenamefont {Koszul}, \citenamefont
  {Fabre},\ and\ \citenamefont {Zimmer}}]{won12}%
  \BibitemOpen
  \bibfield  {author} {\bibinfo {author} {\bibfnamefont {H.}~\bibnamefont
  {Wong}}, \bibinfo {author} {\bibfnamefont {H.}~\bibnamefont {Marie-Nelly}},
  \bibinfo {author} {\bibfnamefont {S.}~\bibnamefont {Herbert}}, \bibinfo
  {author} {\bibfnamefont {P.}~\bibnamefont {Carrivain}}, \bibinfo {author}
  {\bibfnamefont {H.}~\bibnamefont {Blanc}}, \bibinfo {author} {\bibfnamefont
  {R.}~\bibnamefont {Koszul}}, \bibinfo {author} {\bibfnamefont
  {E.}~\bibnamefont {Fabre}},\ and\ \bibinfo {author} {\bibfnamefont
  {C.}~\bibnamefont {Zimmer}},\ }\bibfield  {title} {\bibinfo {title} {A
  predictive computional model of the dynamic 3d interphase yeast nucleus},\
  }\href@noop {} {\bibfield  {journal} {\bibinfo  {journal} {Curr Biol}\
  }\textbf {\bibinfo {volume} {22}},\ \bibinfo {pages} {1881} (\bibinfo {year}
  {2012})}\BibitemShut {NoStop}%
\bibitem [{\citenamefont {Altrock}\ \emph {et~al.}(2011)\citenamefont
  {Altrock}, \citenamefont {Traulsen},\ and\ \citenamefont {Reed}}]{alt11}%
  \BibitemOpen
  \bibfield  {author} {\bibinfo {author} {\bibfnamefont {P.}~\bibnamefont
  {Altrock}}, \bibinfo {author} {\bibfnamefont {A.}~\bibnamefont {Traulsen}},\
  and\ \bibinfo {author} {\bibfnamefont {F.}~\bibnamefont {Reed}},\ }\bibfield
  {title} {\bibinfo {title} {Stability properties of underdominance in finite
  subdivided populations},\ }\href@noop {} {\bibfield  {journal} {\bibinfo
  {journal} {PLoS Computational Bio.}\ }\textbf {\bibinfo {volume} {7}},\
  \bibinfo {pages} {e100260} (\bibinfo {year} {2011})}\BibitemShut {NoStop}%
\bibitem [{\citenamefont {Láruson}\ and\ \citenamefont {Reed}(2016)}]{lar16}%
  \BibitemOpen
  \bibfield  {author} {\bibinfo {author} {\bibfnamefont {A.}~\bibnamefont
  {Láruson}}\ and\ \bibinfo {author} {\bibfnamefont {F.}~\bibnamefont
  {Reed}},\ }\bibfield  {title} {\bibinfo {title} {Stability of underdominant
  genetic polymorphisms in population networks},\ }\href@noop {} {\bibfield
  {journal} {\bibinfo  {journal} {Journal of Theoretical Biology}\ }\textbf
  {\bibinfo {volume} {390}},\ \bibinfo {pages} {156} (\bibinfo {year}
  {2016})}\BibitemShut {NoStop}%
\bibitem [{\citenamefont {Darwin}(1859)}]{dar59}%
  \BibitemOpen
  \bibfield  {author} {\bibinfo {author} {\bibfnamefont {C.}~\bibnamefont
  {Darwin}},\ }\href@noop {} {\emph {\bibinfo {title} {On the Origin of Species
  by Means of Natural Selection, or the Preservation of Favoured Races in the
  Struggle for Life}}}\ (\bibinfo  {publisher} {John Murray},\ \bibinfo {year}
  {1859})\BibitemShut {NoStop}%
\bibitem [{\citenamefont {Thomson}(1962)}]{tom62}%
  \BibitemOpen
  \bibfield  {author} {\bibinfo {author} {\bibfnamefont {W.}~\bibnamefont
  {Thomson}},\ }\bibfield  {title} {\bibinfo {title} {On the age of the sun's
  heat},\ }\href@noop {} {\bibfield  {journal} {\bibinfo  {journal}
  {Macmillan's Magazine}\ }\textbf {\bibinfo {volume} {5}},\ \bibinfo {pages}
  {388} (\bibinfo {year} {1962})}\BibitemShut {NoStop}%
\bibitem [{\citenamefont {SantaLucia}\ and\ \citenamefont
  {Hicks}(2004)}]{san04}%
  \BibitemOpen
  \bibfield  {author} {\bibinfo {author} {\bibfnamefont {J.}~\bibnamefont
  {SantaLucia}}\ and\ \bibinfo {author} {\bibfnamefont {D.}~\bibnamefont
  {Hicks}},\ }\bibfield  {title} {\bibinfo {title} {he thermodynamics of dna
  structural motifs},\ }\href@noop {} {\bibfield  {journal} {\bibinfo
  {journal} {Annu. Rev. Biophys. Biomol. Struct.}\ }\textbf {\bibinfo {volume}
  {133}},\ \bibinfo {pages} {415} (\bibinfo {year} {2004})}\BibitemShut
  {NoStop}%
\bibitem [{\citenamefont {Travers}\ and\ \citenamefont
  {Mushhelishvilli}(2013)}]{tra13}%
  \BibitemOpen
  \bibfield  {author} {\bibinfo {author} {\bibfnamefont {A.}~\bibnamefont
  {Travers}}\ and\ \bibinfo {author} {\bibfnamefont {G.}~\bibnamefont
  {Mushhelishvilli}},\ }\bibfield  {title} {\bibinfo {title} {Dna
  thermodynamics shape chromosome organization and topology},\ }\href@noop {}
  {\bibfield  {journal} {\bibinfo  {journal} {Biochem. Soc. Trans.}\ }\textbf
  {\bibinfo {volume} {41}},\ \bibinfo {pages} {548} (\bibinfo {year}
  {2013})}\BibitemShut {NoStop}%
\bibitem [{\citenamefont {Wang}\ \emph {et~al.}(2016)\citenamefont {Wang},
  \citenamefont {Nocka}, \citenamefont {Wiemann}, \citenamefont {Hinckley},
  \citenamefont {Mukerji},\ and\ \citenamefont {Starr}}]{wan16}%
  \BibitemOpen
  \bibfield  {author} {\bibinfo {author} {\bibfnamefont {W.}~\bibnamefont
  {Wang}}, \bibinfo {author} {\bibfnamefont {L.}~\bibnamefont {Nocka}},
  \bibinfo {author} {\bibfnamefont {B.}~\bibnamefont {Wiemann}}, \bibinfo
  {author} {\bibfnamefont {D.}~\bibnamefont {Hinckley}}, \bibinfo {author}
  {\bibfnamefont {I.}~\bibnamefont {Mukerji}},\ and\ \bibinfo {author}
  {\bibfnamefont {F.}~\bibnamefont {Starr}},\ }\bibfield  {title} {\bibinfo
  {title} {Holliday junction thermodynamics and structure: coarse-grained
  simulations and experiments},\ }\href@noop {} {\bibfield  {journal} {\bibinfo
   {journal} {Sci. Rep.}\ }\textbf {\bibinfo {volume} {6}},\ \bibinfo {pages}
  {doi:10.1038/journal.srep22863} (\bibinfo {year} {2016})}\BibitemShut
  {NoStop}%
\bibitem [{\citenamefont {Connallon}\ and\ \citenamefont
  {Olito}(2021)}]{con21}%
  \BibitemOpen
  \bibfield  {author} {\bibinfo {author} {\bibfnamefont {T.}~\bibnamefont
  {Connallon}}\ and\ \bibinfo {author} {\bibfnamefont {C.}~\bibnamefont
  {Olito}},\ }\bibfield  {title} {\bibinfo {title} {Natural selection and the
  distribution of chromosomal inversion lengths},\ }\href@noop {} {\bibfield
  {journal} {\bibinfo  {journal} {Molecular Ecology}\ ,\ \bibinfo {pages}
  {doi.org/10.1111/mec.16091}} (\bibinfo {year} {2021})}\BibitemShut {NoStop}%
\bibitem [{\citenamefont {Eaton}\ \emph {et~al.}(2019)\citenamefont {Eaton},
  \citenamefont {Bateman}, \citenamefont {Hauberg},\ and\ \citenamefont
  {Wehbring}}]{eat19}%
  \BibitemOpen
  \bibfield  {author} {\bibinfo {author} {\bibfnamefont {J.~W.}\ \bibnamefont
  {Eaton}}, \bibinfo {author} {\bibfnamefont {D.}~\bibnamefont {Bateman}},
  \bibinfo {author} {\bibfnamefont {S.}~\bibnamefont {Hauberg}},\ and\ \bibinfo
  {author} {\bibfnamefont {R.}~\bibnamefont {Wehbring}},\ }\href@noop {} {\emph
  {\bibinfo {title} {{GNU Octave} version 5.1.0 manual: a high-level
  interactive lan guage for numerical computations}}} (\bibinfo {year}
  {2019}),\ \bibinfo {note}
  {https://www.gnu.org/software/octave/doc/v5.1.0/}\BibitemShut {NoStop}%
\bibitem [{\citenamefont {Huang}(1963)}]{hua63}%
  \BibitemOpen
  \bibfield  {author} {\bibinfo {author} {\bibfnamefont {K.}~\bibnamefont
  {Huang}},\ }\href@noop {} {\emph {\bibinfo {title} {Statistical
  Mechanics}}},\ \bibinfo {edition} {1st}\ ed.\ (\bibinfo  {publisher}
  {Wiley},\ \bibinfo {year} {1963})\BibitemShut {NoStop}%
\bibitem [{\citenamefont {Young}\ and\ \citenamefont {Willson}(1987)}]{you87}%
  \BibitemOpen
  \bibfield  {author} {\bibinfo {author} {\bibfnamefont {J.}~\bibnamefont
  {Young}}\ and\ \bibinfo {author} {\bibfnamefont {L.}~\bibnamefont
  {Willson}},\ }\bibfield  {title} {\bibinfo {title} {Use of bose-einstein
  statistics in population dynamics models of arthropods},\ }\href@noop {}
  {\bibfield  {journal} {\bibinfo  {journal} {Ecological Modelling}\ }\textbf
  {\bibinfo {volume} {36}},\ \bibinfo {pages} {89} (\bibinfo {year}
  {1987})}\BibitemShut {NoStop}%
\bibitem [{\citenamefont {Wilson}\ \emph {et~al.}(1987)\citenamefont {Wilson},
  \citenamefont {Young},\ and\ \citenamefont {Folks}}]{wil87}%
  \BibitemOpen
  \bibfield  {author} {\bibinfo {author} {\bibfnamefont {L.}~\bibnamefont
  {Wilson}}, \bibinfo {author} {\bibfnamefont {J.}~\bibnamefont {Young}},\ and\
  \bibinfo {author} {\bibfnamefont {J.}~\bibnamefont {Folks}},\ }\bibfield
  {title} {\bibinfo {title} {A biological application of bose-einstein
  statistics populations},\ }\href@noop {} {\bibfield  {journal} {\bibinfo
  {journal} {Comm. in Stats.-Theory and Methods}\ }\textbf {\bibinfo {volume}
  {16}},\ \bibinfo {pages} {445} (\bibinfo {year} {1987})}\BibitemShut
  {NoStop}%
\bibitem [{\citenamefont {James}(1989)}]{jam89}%
  \BibitemOpen
  \bibfield  {author} {\bibinfo {author} {\bibfnamefont {N.}~\bibnamefont
  {James}},\ }\bibfield  {title} {\bibinfo {title} {The use of bose-einstein
  statistics in analysing the distribution of intracellular organelles: the
  development of a bose-einstein probe},\ }\href@noop {} {\bibfield  {journal}
  {\bibinfo  {journal} {Experientia}\ }\textbf {\bibinfo {volume} {45}},\
  \bibinfo {pages} {1078} (\bibinfo {year} {1989})}\BibitemShut {NoStop}%
\bibitem [{\citenamefont {Strittmatter}\ \emph {et~al.}(1999)\citenamefont
  {Strittmatter}, \citenamefont {Schnier}, \citenamefont {Klassen},\ and\
  \citenamefont {Williams}}]{str99}%
  \BibitemOpen
  \bibfield  {author} {\bibinfo {author} {\bibfnamefont {E.}~\bibnamefont
  {Strittmatter}}, \bibinfo {author} {\bibfnamefont {P.}~\bibnamefont
  {Schnier}}, \bibinfo {author} {\bibfnamefont {J.}~\bibnamefont {Klassen}},\
  and\ \bibinfo {author} {\bibfnamefont {E.}~\bibnamefont {Williams}},\
  }\bibfield  {title} {\bibinfo {title} {Dissociation energies of deoxyribose
  nucleotide dimer anions measured using blackbody infrared radiative
  dissociation},\ }\href@noop {} {\bibfield  {journal} {\bibinfo  {journal} {J
  Am Soc Mass Spectrom.}\ }\textbf {\bibinfo {volume} {10}},\ \bibinfo {pages}
  {1095–1104. doi: 10.1016/S1044} (\bibinfo {year} {1999})}\BibitemShut
  {NoStop}%
\bibitem [{\citenamefont {Tsai}\ \emph {et~al.}(2008)\citenamefont {Tsai},
  \citenamefont {Bensasson}, \citenamefont {Burt},\ and\ \citenamefont
  {Koufopanou}}]{tsa08}%
  \BibitemOpen
  \bibfield  {author} {\bibinfo {author} {\bibfnamefont {I.}~\bibnamefont
  {Tsai}}, \bibinfo {author} {\bibfnamefont {D.}~\bibnamefont {Bensasson}},
  \bibinfo {author} {\bibfnamefont {A.}~\bibnamefont {Burt}},\ and\ \bibinfo
  {author} {\bibfnamefont {V.}~\bibnamefont {Koufopanou}},\ }\bibfield  {title}
  {\bibinfo {title} {Population genomics of teh wild yeast
  \textit{Saccharomyces paradoxus}:quantifying the life cycle},\ }\href@noop {}
  {\bibfield  {journal} {\bibinfo  {journal} {Proc. Natl Acad. Sci.}\ }\textbf
  {\bibinfo {volume} {105}},\ \bibinfo {pages} {4957} (\bibinfo {year}
  {2008})}\BibitemShut {NoStop}%
\bibitem [{\citenamefont {Nieuwenhuis}\ and\ \citenamefont
  {James}(2016)}]{nie16}%
  \BibitemOpen
  \bibfield  {author} {\bibinfo {author} {\bibfnamefont {B.}~\bibnamefont
  {Nieuwenhuis}}\ and\ \bibinfo {author} {\bibfnamefont {T.}~\bibnamefont
  {James}},\ }\bibfield  {title} {\bibinfo {title} {The frequency of sex in
  fungi},\ }\href@noop {} {\bibfield  {journal} {\bibinfo  {journal} {Philos
  Trans R Soc Lond B Biol Sci.}\ }\textbf {\bibinfo {volume} {371}},\ \bibinfo
  {pages} {doi: 10.1098/rstb.2015.0540} (\bibinfo {year} {2016})}\BibitemShut
  {NoStop}%
\bibitem [{\citenamefont {Fay}\ and\ \citenamefont {Benavides}(2005)}]{fay05}%
  \BibitemOpen
  \bibfield  {author} {\bibinfo {author} {\bibfnamefont {J.}~\bibnamefont
  {Fay}}\ and\ \bibinfo {author} {\bibfnamefont {J.~A.}\ \bibnamefont
  {Benavides}},\ }\bibfield  {title} {\bibinfo {title} {Evidence for
  domesticated and wild populations of \textit{Saccharomyces cerevisiae}},\
  }\href@noop {} {\bibfield  {journal} {\bibinfo  {journal} {PLoS Genet.}\
  }\textbf {\bibinfo {volume} {1}},\ \bibinfo {pages}
  {10.1371/journal.pgen.0010005} (\bibinfo {year} {2005})}\BibitemShut
  {NoStop}%
\bibitem [{\citenamefont {Gallone}\ \emph {et~al.}(2016)\citenamefont
  {Gallone}, \citenamefont {Steensels}, \citenamefont {Prahl}, \citenamefont
  {Soriaga}, \citenamefont {Saels}, \citenamefont {Herrera-Malaver},
  \citenamefont {Merlevede}, \citenamefont {Roncoroni}, \citenamefont
  {Voordeckers},\ and\ \citenamefont {et~al.}}]{gal16}%
  \BibitemOpen
  \bibfield  {author} {\bibinfo {author} {\bibfnamefont {B.}~\bibnamefont
  {Gallone}}, \bibinfo {author} {\bibfnamefont {J.}~\bibnamefont {Steensels}},
  \bibinfo {author} {\bibfnamefont {T.}~\bibnamefont {Prahl}}, \bibinfo
  {author} {\bibfnamefont {L.}~\bibnamefont {Soriaga}}, \bibinfo {author}
  {\bibfnamefont {V.}~\bibnamefont {Saels}}, \bibinfo {author} {\bibfnamefont
  {B.}~\bibnamefont {Herrera-Malaver}}, \bibinfo {author} {\bibfnamefont
  {A.}~\bibnamefont {Merlevede}}, \bibinfo {author} {\bibfnamefont
  {M.}~\bibnamefont {Roncoroni}}, \bibinfo {author} {\bibfnamefont
  {K.}~\bibnamefont {Voordeckers}},\ and\ \bibinfo {author} {\bibfnamefont
  {L.~M.}\ \bibnamefont {et~al.}},\ }\bibfield  {title} {\bibinfo {title}
  {Domestication and divergence of \textit{Saccharomyces cervisiae} beer
  yeasts},\ }\href@noop {} {\bibfield  {journal} {\bibinfo  {journal} {Cell}\
  }\textbf {\bibinfo {volume} {166}},\ \bibinfo {pages} {1397} (\bibinfo {year}
  {2016})}\BibitemShut {NoStop}%
\end{thebibliography}%

\end{document}